\documentclass[preprint,12pt,authoryear]{elsarticle}

\usepackage{amssymb}
\usepackage{multirow}
\usepackage{url}
\usepackage{amsmath}

\journal{Elsevier}
\begin{document}

\begin{frontmatter}



\title{Leveraging Connected Vehicle Data for Near-Crash Detection and Analysis in Urban Environments}
\author[label1]{Xinyu Li}
\author[label2]{Dayong (Jason) Wu \corref{cor1}}
\cortext[cor1]{Corresponding author: J-Wu@tti.tamu.edu}
\author[label1,label3]{Xinyue Ye}
\author[label1]{Quan Sun}

\address[label1]{Department of Landscape Architecture and Urban Planning, Texas A\&M University, College Station, Texas, United States, 77840}
            
\address[label2]{Texas A\&M Transportation Institute, 	12700 Park Central Dr., Suite 1000, Dallas, TX 75251}
             
\address[label3]{The Center for Geospatial Sciences, Applications, and Technology, Texas A\&M University, College Station, Texas, United States, 77840}
             




\begin{abstract}
Urban traffic safety is a pressing concern in modern transportation systems, especially in rapidly growing metropolitan areas where increased traffic congestion, complex road networks, and diverse driving behaviors exacerbate the risk of traffic incidents. Traditional traffic crash data analysis offers valuable insights but often overlooks a broader range of road safety risks. Near-crash events, which occur more frequently and signal potential collisions, provide a more comprehensive perspective on traffic safety. However, city-scale analysis of near-crash events remains limited due to the significant challenges in large-scale real-world data collection, processing, and analysis. This study utilizes one month of connected vehicle data, comprising billions of records, to detect and analyze near-crash events across the road network in the City of San Antonio, Texas. We propose an efficient framework integrating spatial-temporal buffering and heading algorithms to accurately identify and map near-crash events. A binary logistic regression model is employed to assess the influence of road geometry, traffic volume, and vehicle types on near-crash risks. Additionally, we examine spatial and temporal patterns, including variations by time of day, day of the week, and road category. The findings of this study show that the vehicles on more than half of road segments will be involved in at least one near-crash event. In addition, more than 50\% near-crash events involved vehicles traveling at speeds over 57.98 mph, and many occurred at short distances between vehicles. The analysis also found that wider roadbeds and multiple lanes reduced near-crash risks, while single-unit trucks slightly increased the likelihood of near-crash events. Finally, the spatial-temporal analysis revealed that near-crash risks were most prominent during weekday peak hours, especially in downtown areas. The findings provide valuable insights for targeted safety interventions and contribute to more effective road safety strategies aimed at reducing serious traffic incidents in urban environments.
\end{abstract}



\begin{keyword}
Urban traffic safety \sep Near-crash event \sep Connected vehicle data \sep Spatial-temporal analysis \sep Road safety interventions


\end{keyword}

\end{frontmatter}



\section{Introduction}
\label{sec1}
Traffic safety and human mobility remain a pressing concern worldwide, particularly in rapidly expanding metropolitan areas where increasing traffic congestion, complex road networks, extreme weather, and diverse road user behaviors converge to heighten the risk of traffic incidents \citep{evans1991traffic,evans2004traffic,ewing2009built,li2024beyond,shinar2017traffic,ye2024enhancing}. Traffic accidents are a pervasive and significant issue in daily urban life, resulting not only in substantial economic losses and infrastructure damage but also in severe injuries and fatalities \citep{gopalakrishnan2012public, zhu2024equity}. In 2023, over 1 million lives were lost globally due to traffic accidents, with countless more affected by non-fatal injuries, underscoring the urgency of addressing road safety as a critical public health challenge in urban environments \citep{who2023road}. The frequency and severity of these incidents make it imperative to develop more effective strategies to mitigate risks and enhance the safety of urban roadways. Traditional traffic safety research has primarily focused on analyzing crash data to understand and mitigate these risks \citep{lord2010statistical,park2009application}. However, crashes, while critical indicators of road safety issues represent only the tip of the iceberg. They occur relatively infrequently, making it difficult to capture the full spectrum of daily risky driving behaviors. 

Near-crash events, characterized by close incidents that could have resulted in a collision, are far more common than actual accidents in our daily lives \citep{perkins1968traffic,steele1999occupational,talebpour2014near}. They provide a crucial window into the precursors of serious accidents, but they are often overlooked in traffic safety studies due to the challenges in detecting and analyzing such events. Leveraging near-crash events facilitates understanding valuable insights into the driving conditions, road characteristics, and behaviors that increase the likelihood of a crash \citep{klauer2006impact, wang2015driving,bakhit2018crash}. By studying near-crash events, researchers can identify spatial-temporal patterns and risk factors that are not evident in crash data alone, enabling the development of more proactive and preventative traffic safety measures.

Recent advancements in connected vehicle technology offer unprecedented opportunities to collect high-frequency, precise positioning data on vehicle movements and interactions. The real-time connected vehicle data enables the detection and analysis of near-crash events in the real world, providing a more comprehensive view of road safety risks \citep{papadoulis2019evaluating,islam2023traffic,islam2023understanding}. Many scholars contribute their great efforts to analyze driving behaviors and traffic safety using connected vehicle data. Many improvements in traffic safety analysis have been achieved \citep{zhang2022real, zheng2021modeling,nazir2023car,ali2023assessing}. However, there remains a gap in the literature regarding the application of connected vehicle data to uncover spatial-temporal patterns of near-crash events, particularly in large-scale urban environments. Few studies systematically investigated the characteristics of the spatial and temporal distribution of near-crash events across different roadway types at the large-scale city level. Understanding these patterns is crucial for developing targeted interventions that can mitigate risks and enhance urban road safety, ultimately reducing the probability of serious accidents.

To fill this gap in urban traffic safety, this study leverages Wejo connected vehicle data (encompassing billions of vehicle records) to detect and map near-crash events, providing a granular view of where and when these incidents occur across the City of San Antonio, Texas. Firstly, we propose a near-crash event detection framework with spatial-temporal buffering and the heading algorithm to accurately identify and spatially map near-crash events across the city’s diverse road network. Based on the detected events, we adopt a binary logic regression model to investigate the road and traffic characteristics, such as road geometry and traffic volume, contributing to higher near-crash risks. Then, we further explore the spatial and temporal distributions of near-crash events, identifying patterns that vary by time of day, day of the week, and specific road types. By examining the distribution on both workdays and holidays (including the weekends and public holidays), the study aims to reveal how hourly and daily traffic patterns influence the probability of near-crash incidents. Through the analysis of near-crash events, this paper aims to answer the following three research questions:
\begin{enumerate}
  \item How can connected vehicle data be utilized to detect and map near-crash events across a diverse urban roadway network with low computational costs?
  \item Which road facilities and traffic characteristics contribute most significantly to the likelihood of near-crash events?
  \item What are the spatial and temporal patterns of near-crash events in San Antonio, and how do these patterns vary across different times of day, days of the week, and road categories?
\end{enumerate}

This study's contributions are grounded in addressing the research questions and achieving the set objectives, thereby advancing the field of traffic safety analysis. First, this study introduces an efficient data-driven approach for detecting near-crash events using a massive amount of connected vehicle data in a metropolitan area. A large-scale analysis enables transportation authorities and urban planners to identify specific locations that require safety interventions, thus contributing to the prevention of future crashes. Second, this study offers insights into the road facilities and traffic characteristics that increase or mitigate near-crash risks. The systematic analysis in this research identifies critical factors such as road geometry, Annual Average Daily Traffic volume (AADT), and vehicle types in AADT that significantly influence the likelihood of near-crash events. By understanding these key contributing factors, policymakers can develop more effective strategies to reduce the occurrence of near-crash events and improve overall traffic safety. The last one is to provide a systematic spatial-temporal analysis of near-crash events, which provides a deeper understanding of the dynamic nature of road safety risks. By mapping these spatial-temporal variations, the study highlights the importance of considering temporal factors when designing and implementing safety measures

The rest of this paper is organized as follows. Section \ref{sec:related_work} reviews and summarizes the relevant studies on road traffic safety, near-crash analysis using connected vehicle data, and related analysis approaches. The overall descriptions of data and study area are provided in Section \ref{sec:dataset}. The framework of near-crash event detection and the models used for regression and spatial-temporal analysis are introduced in Section \ref{sec:method}. In Section \ref{sec:results}, the analysis results are illustrated. Finally, we conclude this study and give a comprehensive discussion in Section \ref{sec:conclusion}.

\section{Literature Review}\label{sec:related_work}
Road traffic safety is a critical and widely studied topic in modern transportation systems, as it plays a fundamental role in shaping public health outcomes, economic resilience, and the overall quality of life in both urban and rural contexts. Specifically, road traffic crashes have emerged as one of the most serious threats to the safety of road users, including pedestrians, cyclists, motorists, and vehicle passengers. Thus, numerous scholars have contributed their efforts to investigate the factors that lead to road traffic crashes from the perspectives of road geometric design, driving performance, and traffic interventions \citep{deery1999young,arvin2019role,suriyawongpaisal2003road,miaou1993modeling}. Their works provided more comprehensive and valuable insights into reducing the risks of road traffic crashes in real-world scenarios. For example, the authors find that regions with infrequency geometric inconsistency could demonstrate higher crash risks, while the areas with similar geometric inconsistency represent lower crash ratios \citep{shilpa2024incorporating}. Also, improvements in roadway width, pedestrian facilities, and access management can efficiently enhance traffic safety \citep{world2023pedestrian,ben2011effect}. However, the majority of these studies rely on historical crash records, which only capture a limited portion of the actual risk landscape. This reliance can constrain the ability to proactively identify and address potential hazards. Furthermore, crash records often suffer from biases, inaccuracies, or missing information regarding location, time, and other relevant attributes, which can lead to misleading findings and conclusions \citep{xie2019use,ali2023assessing,islam2023traffic}.

The concept of near-crash events as a surrogate measure for traffic safety was first introduced by \citep{perkins1968traffic} and has since evolved with advancements in data collection technologies. At the early stage, the near-crash events are identified based on traditional observational approaches or basic Traffic Conflict Techniques \citep{parker1989traffic}, which rely on manual observation and interpretation of driver behaviors and vehicle interactions\citep{hyden1987development, steele1999occupational}. With mobile and low-cost sensors emerging, naturalistic driving data are used to detect and analyze near-crash events. The naturalistic driving data is collected from a series of sensors, which documents the driving states (e.g., vehicle speed, acceleration, and braking), driver's control, lane position, and frequency of behaviors in normal driving. Based on the naturalistic driving data, studies find that the velocity when braking, lane-changing behavior on highways, and yaw rate of vehicles highly contribute to near-crash events \citep{kong2021mining,perez2017performance,wang2015driving}. Moreover, distraction or inattention when driving can lead to high crash or near-crash risks, such as drowsy driving, texting with a mobile phone, and slow eye movement during night driving \citep{lee2016high,klauer2006impact,crummy2008prevalence,nasr2021prevalence}. Although naturalistic driving data can provide wider implications for investigating the factors contributing to crashes and near-crashes, a major challenge lies in the limited sample size and potential bias inherent in this type of data. These limitations may not fully capture the diverse range of driving behaviors and conditions that contribute to near-crash events, necessitating careful consideration in the analysis and interpretation of the results. Moreover, because the volunteers in naturalistic driving data collection are aware that their performance is being monitored, this awareness can lead to altered driving behavior, often referred to as the Hawthorne effect\citep{adair1984hawthorne,golob1998projecting,reagan2013using}. This effect can result in more cautious or atypical driving patterns, which may not accurately reflect their usual behavior, thereby introducing bias and limiting the validity of the data collected in naturalistic driving studies.

The emergence of connected vehicle technology has revolutionized near-crash analysis by enabling the collection of high-frequency, high-precise location data on vehicle movements and interactions. This data provides real-time insights into the driving environment, allowing for the detection and analysis of near-crash events as they happen. Studies have demonstrated that connected vehicle data have been used to detect near-crash events and develop near-crash prediction models that identify high-risk situations before the vehicles are involved in the crashes on highways or specific small areas  \citep{islam2023understanding,islam2023traffic,xie2019use,zheng2021modeling}. Moreover, due to the large volume of real-time connected vehicle data that becomes more available, this data provides alternatives for identifying crashes and near-crashes in the upcoming era of connected and autonomous vehicles \citep{papadoulis2019evaluating,virdi2019safety,mannering2020big}. This capability will enable more proactive and precise road safety interventions, potentially reducing the occurrence of traffic incidents and enhancing overall traffic safety. Besides, connected vehicle data is more widely used in urban traffic flow, speed estimation, and microscope traffic simulation applications \citep{kamal2018road,sun2016integrated,gueriau2016assess,lu2020impact}. 

However, few studies concentrate on analyzing spatial-temporal patterns of near-crash events in large-scale or city-level scenarios. Understanding these patterns is crucial for identifying high-risk areas and times, which can inform the development of targeted safety interventions. Most existing research has been limited to smaller-scale studies or specific road segments, leaving a significant gap in our knowledge of how near-crash events are distributed across entire urban road networks. Analyzing these patterns at a city level allows for a more comprehensive understanding of the underlying factors contributing to near-crash events, such as traffic density, road design, and varying driver behaviors during different time periods of the day or week. This broader perspective is essential for developing effective strategies to mitigate risks and enhance road safety in increasingly complex urban environments.

\section{Dataset and Study Area}
\label{sec:dataset}
This study uses the Wejo connected vehicle dataset to detect near-crash events. The Wejo dataset is collected from mainstream vehicle manufacturers in the United States. It primarily documents non-commercial fleet movement data, which can better reflect the driving situation and urban traffic states on the roadway. The telematics devices installed on the vehicles send instantaneous information to the Wejo Cloud Platform in near-real time. This dataset documents the vehicle's states from the engine on to off, including GPS location as latitude ($\varphi$) and longitude ($\lambda$), speed, heading, journey ID, data point ID, and timestamp of data documented. Table \ref{table:wejodata} shows example records of Wejo Dataset. The Wejo dataset provides meter-level positioning accuracy (6 decimal places) and high-frequency sampling interval (every 3 seconds), facilitating near-crash event detection on the roadway. The one-month Wejo data from 11/01/2021 to 11/30/2021 are used in this study. The raw dataset documents over 2.9 billion GPS data points from 7.9 million unique trajectories this month. We adopt preprocessing methods to remove the outliers caused by GPS signal drifts in the raw data to keep each trajectory smooth and reasonably distributed on the roadway. For the data preprocessing and following analysis, we adopt parallel and distributed computation approaches to handle such a huge volume of data. 
\begin{table}[htbp]
    \caption{Example Records of Wejo Connected Vehicle Data in 11/01/2021}\label{table:wejodata}
        \resizebox{\textwidth}{!}{%
        \begin{tabular}{cccccccc}
            \hline
            \textbf{Data Point ID} & \textbf{Journey ID} & \textbf{Timestamp ($t$)} & \textbf{Lat($\varphi$)} & \textbf{Lon($\lambda$)} & \textbf{Speed} & \textbf{Heading ($\vartheta$)} & \textbf{Ignition Status} \\\hline
            105623589 & 94a69188fc92e9bde & 08:10:18 & 29.****** & -98.****** & 0 & 238 & KEY ON \\
            236598742 & 4f9cba35e35ac567d & 15:21:39 & 29.****** & -98.****** & 25 & 254 & MID JOURNEY \\
            \multirow{2}{*}{\vdots} & & \multirow{2}{*}{\vdots} & \multirow{2}{*}{\vdots} & \multirow{2}{*}{\vdots} & & & \multirow{2}{*}{\vdots} \\
            & & & & & & & \\
            453219856 & 94a69188fc92e9bde & 08:15:59 & 29.****** & -98.****** & 45 & 32 & MID JOURNEY \\
            105623511 & 785e1b0b6c68d3069 & 07:10:21 & 29.****** & -98.****** & 50 & 185 & MID JOURNEY \\
            598746523 & 4f9cba35e35ac567d & 15:25:29 & 29.****** & -98.****** & 60 & 25 & MID JOURNEY \\
            263598421 & 785e1b0b6c68d3069 & 07:15:35 & 29.****** & -98.****** & 0 & 228 & KEY OFF \\\hline
        \end{tabular}
        }
\end{table}

\begin{figure}[!ht]
\centering
\includegraphics[width=1.0\textwidth]{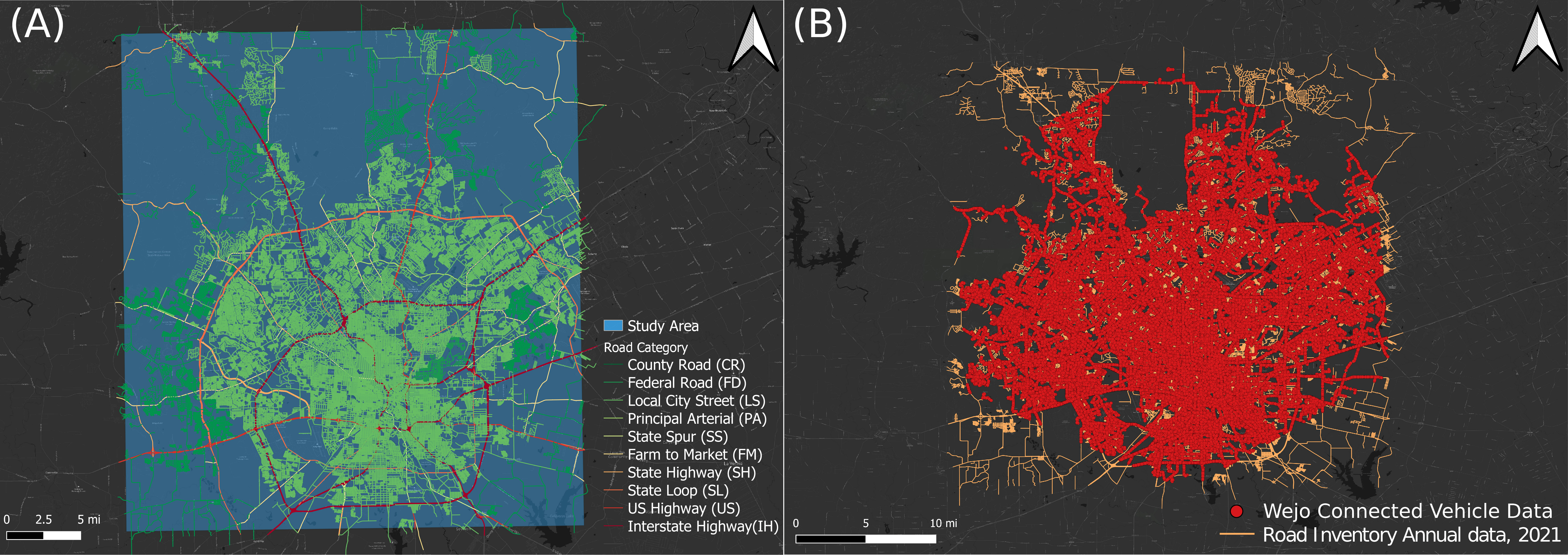}
\caption{Study area in the City of San Antonio. (A) represents the road network data in San Antonio, and (B) represents the spatial distribution of Wejo connected vehicle data records on 11/01/2021.}
\label{fig:study_area}
\end{figure}

As shown in Fig.\ref{fig:study_area}, the study area covers the majority of the city of San Antonio, Texas, across 70 zip code areas (close to 1660.95 square miles). Over 1.45 million people are living in this city. Based on the travel survey in San Antonio, over 60$\%$ surveyed households use private vehicles as the travel model in their daily life. Understanding the ratio of near-crash events on road segments can significantly improve traffic safety for the traffic department. This study uses the annual data from the road inventory provided by the Texas Department of Transportation (TxDOT)\footnote{TxDOT Data Portal: \url{https://www.txdot.gov/data-maps/roadway-inventory.html}.}. The road inventory dataset has ten road categories, including three types of local roads and seven types of highways maintained by the TxDOT. The detected near-crash events will fall under  13,231 road segments in the TxDOT road inventory data. The detailed attributes of each road segment are provided in \ref{app1:table}. The following analysis will adopt these attributes in binary logistic regression to reveal the road facilities that could increase or lower the probability of near-crash events.  

\section{Methodology}
\label{sec:method}
In this section, we develop an algorithm framework that leverages the connected vehicles' trajectories to detect near-crash events, as illustrated in Fig.\ref{fig:fig1}. The near-crash detection algorithm leverages the heading angles and current locations of two vehicles to estimate their potential conflict location. We can further calculate the time of two vehicles reaching the conflict location based on their current speeds. When the difference between the two vehicles' reaching time is less than the Time-To-Collision (TTC) threshold (3 seconds), we assume these two vehicles are involved in a near-crash event \citep{islam2023understanding, islam2023traffic, li2017evaluation,xing2019examining}. To decrease the computational costs of near-crash event detection, we create a spatial-temporal buffer for data points of each trajectory to query the data points of other vehicles that are close in space and time. After near-crash event detection, we further match the events on the road segments. This facilitates investigating the correlation between traffic facilities and near-trash events on the roadway and revealing the spatial-temporal patterns of near-crash events in San Antonio. We adopt a binary logistic regression model and the $Getis$-$Ord Gi^*$ algorithm to examine the correlation and uncover the spatial-temporal patterns, respectively \citep{sarkar2011logistic,getis1992analysis}. 

\subsection{Near-Crash Event Detection}
\begin{figure}[htbp]
\centering
\includegraphics[width=1.0\textwidth]{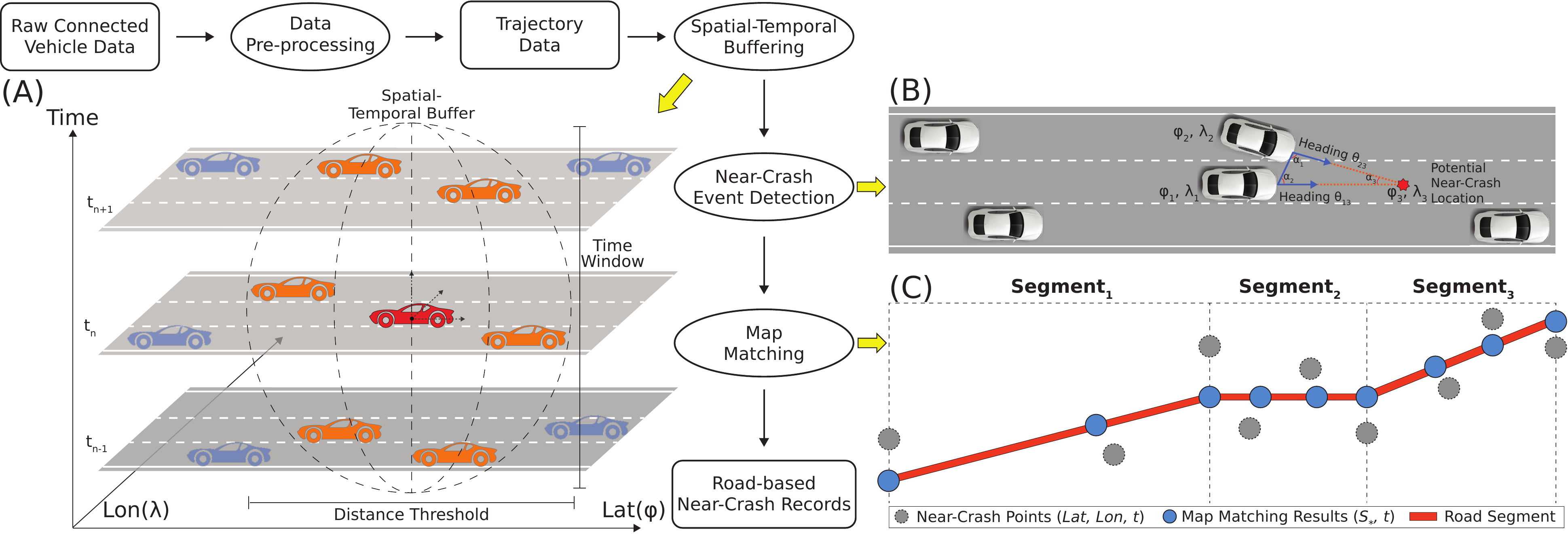}
\caption{Framework of Near-Crash Event Detection}\label{fig:fig1}
\end{figure}

As shown in Fig.\ref{fig:fig1}, the near-crash event detection framework proposed in this study has three main steps: spatial-temporal buffering, near-crash event detection using the heading algorithm, and a map-matching approach. The raw connected vehicle data is preprocessed and reorganized as the trajectory data. We first create spatial-temporal buffers for each point in the trajectory to identify the data points of other vehicles at a distance of less than 100 meters(distance threshold) and the difference of recorded timestamps less than 10 seconds(time window). The parameters for spatial-temporal buffer settings ensure the identification of vehicles that are in close proximity and have the potential to interact within a short time frame, capturing realistic near-crash scenarios. The data points in the spatial-temporal buffer will be further input to the near-crash event detection algorithm. When the Time-To-Collision (TTC) of a potential near-crash is less than 3 seconds, it will be considered a near-crash event \citep{islam2023understanding, islam2023traffic, li2017evaluation,xing2019examining}. The detected near-crash events will be further matched on the roadway using spatial querying. Finally, the number of near-crash events on each road segment can be calculated.

The first step is to create spatial-temporal buffers for each data point in every trajectory to decrease the computational time complexity of near-crash detection (shown in Fig.\ref{fig:fig1}(A)). The data points in each trajectory will find other spatially and temporally close vehicles' data points through the spatial-temporal buffering approach. We adopt the k-dimensional tree algorithm (KDTree) to query the data points within a distance threshold \citep{maneewongvatana1999analysis, he2012computing}. The KDTree is a type of binary tree where each node represents an axis-aligned rectangle in the space. Each node is associated with a specific axis and divides the points into two groups based on whether their coordinate along that axis is greater than or less than a particular threshold value \citep{maneewongvatana1999analysis}. Given a specific data point $p(\varphi,\lambda,t,\vartheta,v)$ as an anchor point, the KDTree algorithm can efficiently query the data points near the anchor point for a specific distance threshold (100 meters), where $\varphi$ and $\lambda$ denotes the latitude and longitude of the vehicle at timestamp $t$, $\vartheta$ and $v$ represent the heading angle and speed of the vehicle when data point recorded, respectively. Meanwhile, in the spatial buffer, a time window is employed to filter out the data points that are more than 10 seconds before or after the timestamp of the anchor point is documented. In other words, we select data points from other trajectories that are within 100 meters of the anchor point and recorded within ten seconds before or after the anchor point's timestamp for the near-crash event detection. Moreover, the spatial-temporal algorithm uses flags to record the pairs of data points detected in a spatial-temporal buffer to avoid duplicate querying, which can further increase the efficiency of near-crash event detection. 

As shown in Fig.\ref{fig:fig1}(B), the second step of the near-crash detection algorithm is to calculate the potential conflict locations in spatial-temporal buffers and find the conflict pairs of near-crash events \citep{islam2023traffic}. Given a pair of trajectory points from two vehicles $p_1(\varphi_1, \lambda_1, t_1,\vartheta_1,v_1)$ and $p_2(\varphi_2, \lambda_2, t_2,\vartheta_2,v_2)$, we use an intersecting radials algorithm to estimate the intersection location $p_3(\varphi_3, \lambda_3)$ of the two vehicles. Based on the intersecting radials algorithm provided in \citep{williams_aviation_2008}, we use the following equations to calculate the intersection location $\varphi_3, \lambda_3$. 
\begin{equation}
    d_{13} = arctan(\frac{sind_{12} * sin\alpha_1 * sin\alpha_2}{cos\alpha_2 + cos\alpha_1 * cos\alpha_3})
    \label{equ:equ1}
\end{equation}

\begin{equation}
    \varphi_3=arcsin(sin\varphi_1*cos d_{13}+cos\varphi_1*sin d_{13}*cos\vartheta_{13})
    \label{equ:equ2}
\end{equation}

\begin{equation}
    \Delta \lambda = arctan(\frac{sin\vartheta_{13} * sin d_{13} * cos \varphi_1}{cos d_{13} - sin\varphi_1 * sin\varphi_3})
    \label{equ:equ3}
\end{equation}
\begin{equation}
    \lambda_3=(\lambda_1 - \Delta \lambda + \pi) \% (2 * \pi) -\pi
    \label{equ:equ4}
\end{equation}
where $\varphi_3$ and $\lambda_3$ denote the latitude and longitude of the intersection location of two trajectories, respectively. $d_{13}$ represents the distance between $p_1$ and $p_3$, which are calculated based on angle $\alpha_1$ (between the $\boldsymbol{v_{12}}$ and $\boldsymbol{v_{13}}$), angle $\alpha_2$ (between the $\boldsymbol{v_{21}}$ and $\boldsymbol{v_{23}}$), angle $\alpha_3$ (between the $\boldsymbol{v_{32}}$ and $\boldsymbol{v_{31}}$), and distance between $p_1$ and $p_2$. $\alpha_1$, $\alpha_2$, and $\alpha_3$ can be calculated based on the heading angles of the two vehicles. The operator $\%$ is to calculate the modulus after division. All the latitudes and longitudes used in the equations are converted to radians for the calculation. Further details can be found in \citep{williams_aviation_2008}. 

After conflict location estimation, we adopt equations \ref{equ:travel_time} and \ref{equ:ttc_cal} to calculate the Time to Conflict (TTC) of the two vehicles\citep{islam2023understanding, islam2023traffic, li2017integrated}. Inspired by the existing studies focusing on driving behavior and near-crash events analysis, they adopted 1-4 seconds as the TTC threshold to identify the conflict. Thus, we use 3 seconds as the TTC threshold in this study. The pairs of data points whose TTC value is smaller than 3 seconds are identified as near-crash events. 

\begin{equation}
    t_{13} = \frac{d_{13}}{v_1}, t_{23} = \frac{d_{23}}{v_2}
    \label{equ:travel_time}
\end{equation}

\begin{equation}
    TTC = min(t_{13}, t_{23}), if\ t_{13}-1.5s \leq t_{23} \leq t_{13}+1.5s
    \label{equ:ttc_cal}
\end{equation}
where $t_{13}$ and $t_{23}$ denote the time duration of $vehicle_1$ reaching estimated conflict location, and the duration of $vehicle_2$ to estimated conflict location, respectively. $d_{13}$ and $d_{23}$ represent the distance between $p_1$ and estimated conflict location, and the distance between $p_2$ and estimated conflict location, respectively. $v_1$ and $v_2$ denote the speed of $p_1$ and $p_2$. 

The final step of the near-crash event detection approach, shown in Fig.\ref{fig:fig1}(C), is to match the detected near-crash events on the road segments, which can summarize the risks of near-crash events on segments. We adopt the spatial querying approach to join the near-crash events with the nearest road segment. To reduce the impact of incorrect matches on the following analysis, this study only considers near-crash events occurring on the same road direction due to overtaking or failure to decelerate. Note that the vehicle trajectory data does not provide elevation information, which can lead to near-crash events detected on different levels of roads (e.g., flyovers or bridges). After matching the events on the road segments, we can perform further analysis, such as aggregating the frequency of near-crash events by segment and assessing the spatial distribution of these events across the City of San Antonio. This helps in identifying high-risk areas, allowing for targeted interventions to improve road safety. The analysis can be enhanced by incorporating additional attributes, such as traffic volume, road geometry, or environmental factors, to better understand the underlying causes of near-crash events.

\subsection{Binary Logistic Regression Model}
To explore the correlation between the risk of near-crash events on road segments and various road environmental factors as well as additional attributes of transportation system, we employ the Binary Logistic Regression (BLR) model \citep{sarkar2011logistic,sze2014likelihood,bham2012multinomial}. This statistical model is particularly well-suited for our analysis as it allows us to model a categorical dependent variable with more than two possible outcomes corresponding to our study's risk levels associated with near-crash events. The BLR model predicts the probabilities of the different possible outcomes of a categorical dependent variable given a set of independent variables. 

In our study, the dependent variable represents the risk level of a near-crash event occurring on a specific road segment, which can take two discrete categories, including low-risk and high-risk segments. The road risk level is determined by the ratio of the number of near-crash events detected to the total number of vehicles passing through the same road segment, as shown in Eq.\ref{equ:ratio}. The road segments whose near-crash events ratio is smaller than 1\% are identified as low-risk segments, and other segments with ratios greater than 1\% are considered relatively high-risk segments. The purpose of categorization is to ensure that the sample sizes for different categories are relatively balanced, thereby avoiding the bias that may arise from class imbalance. 
\begin{equation}
    Risk\ Ratio_i = \frac{No.\ of\ Near\text{-}Crash\ Events\ on\ Segment_i}{No.\ of\ Vehicles\ Passing\ through\ Segment_i}
    \label{equ:ratio}
\end{equation}

The BLR model predicts the probability $P(Y=1|X)$ that the dependent variable $Y$ is in the high-risk category (corresponding to the low-risk segment, which is $Y=0$) based on the independent variables $X=[X_1, X_2, ..., X_n]$. The logistic regression model is formulated as Eq.\ref{equ:logistic}. The maximum likelihood estimation is adopted to estimate the bias and coefficients in the logistic regression model. The likelihood function $L(\beta)$ is the product of the probabilities of observing dependent values $Y_i$ given the independent variables $X_i$. The likelihood estimation is represented in Eq.\ref{equ:likelihood}. After likelihood estimation, we use log-likelihood and Pesudo $R^2$ to evaluate the regression performance.

\begin{equation}
    log(\frac{P(Y=1|X)}{P(Y=0|X)})=\beta_0+\beta_1X_1+\beta_2X_2+\cdots +\beta_nX_n
    \label{equ:logistic}
\end{equation}
\begin{equation}
    L(\beta)=\prod_{i=1}^{n}P(Y_i|X_i)^{Y_i}(1-P(Y_i|X_i)^{1-Y_i}))
    \label{equ:likelihood}
\end{equation}
where $\beta_0$ is the bias and $\beta_i$ are the coefficients corresponding to each independent variable $X_i$. $n$ is the number of observations, and $P(Y_i|X_i)$ is the probability of the observed value for each observation. 

\subsection{Hot Spot Analysis}
We adopt hot spot analysis in this study to uncover the spatial patterns of near-crash events across the City of San Antonio, allowing for the identification of locations with significantly higher or lower concentrations. By employing spatial statistical approaches, the $Getis$-$Ord Gi^*$ model \citep{manepalli2011evaluation,getis1992analysis,ord1995local} can detect clusters of road segments where near-crash events are more prevalent, known as hot spots, as well as areas with fewer incidents, known as cold spots. The identification of these spatial patterns is crucial for understanding the geographic distribution of road safety risks and for guiding targeted interventions aimed at reducing the frequency and severity of near-crash events in high-risk areas.

The $Getis$-$Ord Gi^*$ model is a typical spatial statistical model in Geographic Information Science (GIS), which adopts $Gi^*$ statistic for each road segment $i$. The calculation of $Gi^*$ is shown in Eq.\ref{equ:hot}, where $x_j$ denotes the ratio of near-crash events at road segment $j$ and $n$ denotes the total number of all road segments. $w_{ij}$ represents the spatial weight between road segments $i$ and $j$, which indicates the influence of $j$ on $i$. $\overline{X}$ and $S$ denote the mean and standard deviation of the near-crash ratio across all road segments, respectively. 
\begin{equation}
    G_i^*=\frac{\sum_{j=1}^{n}w_{ij}x_j-\overline{X}\sum_{j=1}^{n}w_{ij}}{S\sqrt{\frac{\sum_{j=1}^{n}w_{ij}^2-(\sum_{j=1}^{n}w_{ij})^2}{n-1}}}
    \label{equ:hot}
\end{equation}

We use a distance-based approach to identify the spatial weight between road segments. In particular, we adopt $w_{ij}=\frac{1}{d_{ij}}$ to identify the spatial weight between location $i$ and $j$, which implies that the spatial relationships between the two locations decrease when their distance increases \citep{tobler1970computer}. In this study, we adopt a fixed distance (1 mile) as the threshold to calculate the spatial weight between two locations. After calculating $G_i^*$, the high positive value of $G_i^*$ indicates that the location $i$ and its neighbors have higher near-crash risk than expected by random choice. The low negative value means the location and neighboring areas have lower near-crash risk than expected. 

\section{Results}\label{sec:results}
\subsection{Overall distribution of near-crash events}
\begin{figure}[htbp]
\centering
\includegraphics[width=1.0\textwidth]{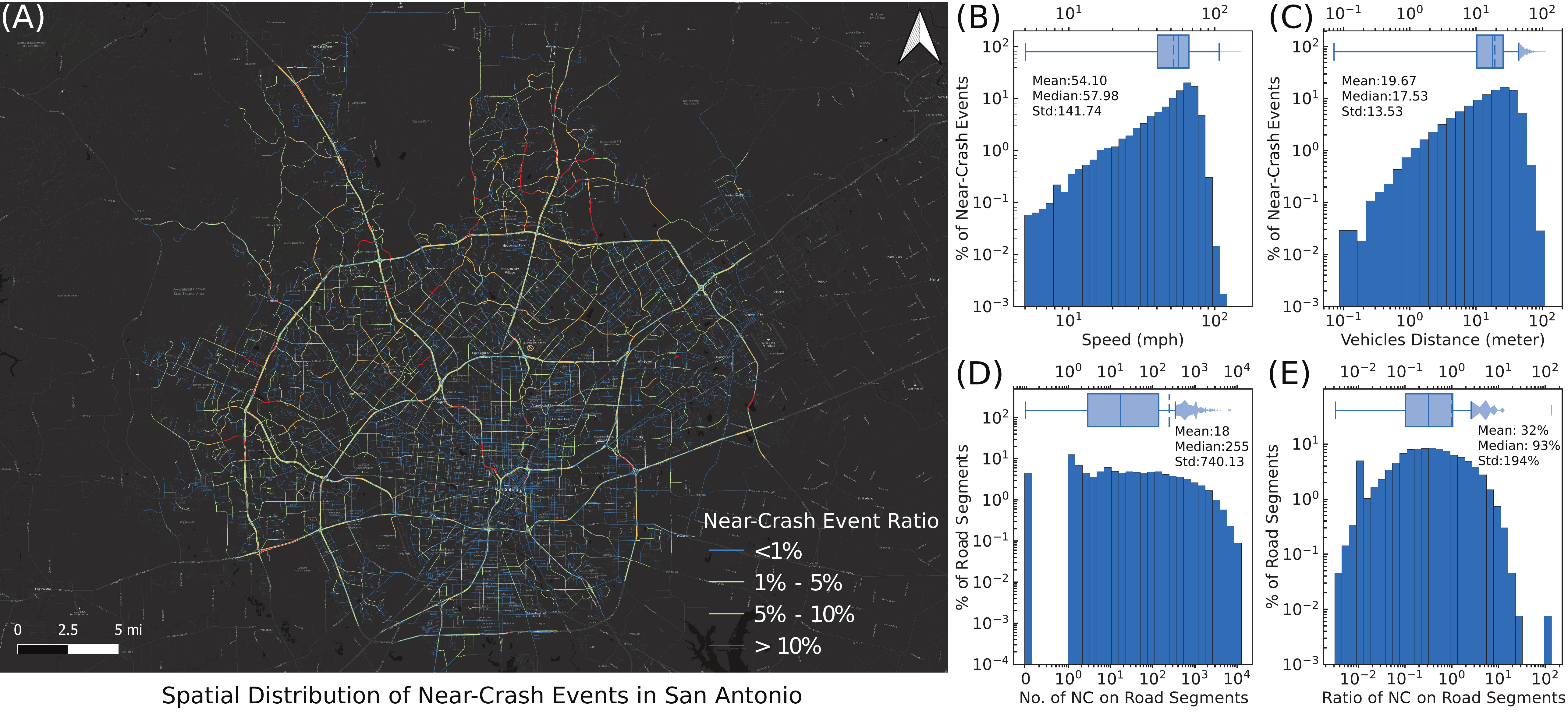}
\caption{Spatial Distribution of Near-Crash Risk Ratio and Attributes of Near-Crash Events in the City of San Antonio}\label{fig:overall}
\end{figure}

The analysis of near-crash events across San Antonio provides critical insights into the spatial distribution and contributing factors of road safety risks. By applying the near-crash events detection approach, 3,377,786 near-crash events are detected from the Wejo connected vehicle dataset across 42 roads with 13,231 road segments in November 2021. Fig.\ref{fig:overall} shows the ratio of near-crash events on road segments during this month and corresponding near-crash attributes in San Antonio. Fig.\ref{fig:overall}(A) demonstrates that near-crash events are not evenly or randomly distributed but are concentrated in certain areas, indicating some urban areas with a high frequency of near-crash incidents. The color gradient in this figure reflects the ratio of near-crash events relative to the observed traffic volume, with darker red shades representing higher risk segments where the near-crash event ratio exceeds 10\%. In contrast, the blue lines indicate lower-risk segments with ratios below 1\%. Notably, there are 9,975 road segments with a ratio of less than 1\%, 2,842 segments with ratios between 1\% and 5\%, and 76 segments where the near-crash event ratio is greater than 10\%. These 76 segments represent potential high-risk areas for near-crash events in the City of San Antonio. 

At the driving behavior level, Fig.\ref{fig:overall}(B) and (C) provide the speed and distance distribution of vehicles involved in near-crash events, which can further offer insights into the vehicles' states when the events occur. Fig.\ref{fig:overall}(B) reflects that the speed of vehicles is over 57.98 mph across over 50\% near-crash events, which indicates that higher speeds are often associated with greater risk. Meanwhile, there is a considerable number of near-crash incidents at low speeds, possibly in more congested or complex urban environments, which requires particular attention for urban transportation planning. Moreover, Fig.\ref{fig:overall}(C) demonstrates the high frequency of near-crash events at shorter distances between two vehicles (smaller than 20m), suggesting that drivers may follow too closely, especially in traffic-dense areas. Thus, enhancing public awareness about safe following distances could help reduce near-crash incidents. 

At the road segment level, Fig.\ref{fig:overall}(D) and (E) reflect the distribution of near-crash events in San Antonio. As shown in Fig.\ref{fig:overall}(D), although many road segments experience a low number of near-crash events, there is a long tail of segments with a significantly higher number of near-crash incidents. The median number of near-crash events on segments is over 255, which indicates that at least half of the road segments analyzed experience a high frequency of near-crash incidents, suggesting that these safety risks are prevalent and not limited to a few isolated locations. Moreover, as shown in Fig.\ref{fig:overall}(E), more than 50\% road segments have a risk ratio greater than 93\%, indicating that almost every vehicle passing over the segment will involve one near-crash event. This suggests the probability of being involved in a traffic accident during daily driving is relatively high in the City of San Antonio. In the following sections, we will further analyze the risk of near-crash events during different time periods on workdays and holidays to reveal the variations in near-crash risk across different urban areas.

Additionally, we adopt the binary logistic regression model to investigate the factors that may affect the risks of near-crash events on road segments. The regression model is conducted by categorizing road segments based on the near-crash event ratio into two groups: low-risk (less than 1\%) and relatively high-risk (greater than 1\%). Also, we adopt the normalization method for all numeric features and use the one-hot encoder for categorical variables in the regression. The low-risk road segments are used as the reference category for the regression model, which means that all the estimated coefficients represent the change in the log odds of a road segment being classified as high-risk compared to the baseline of being low-risk. Moreover, we only select the road segments maintained by TxDOT for the regression model because the attributes of these segments are provided. Some local road segments lack the necessary attribute information for inclusion in the regression. In particular, 2,811,285 near-crash events from 6,832 road segments across 39 roads are considered in the regression model, including 4,750 segments in the low-risk group and 2,082 in the high-risk group. 

After filtering out the covariant and insignificant independent variables, Table \ref{table:logitresult} presents the overall confidence level of the regression model, along with the coefficients for each independent variable. The regression model is highly statistically significant overall based on the $Chi-square$ and $P$-value indicators. Although the Pseudo $R^2$ is 0.1217, indicating that the model explains a modest portion of the variance in near-crash risk, the statistically significant coefficients suggest that certain road segments and traffic attributes still have a meaningful impact on the likelihood of a road segment being classified as high-risk. The possible reason for the low Pseudo $R^2$ is that near-crash events are heavily influenced by driver behavior and driving conditions, which are not fully captured by the road and traffic attributes included in the model. 

\begin{table}[htbp]
\caption{Binary Logistic Regression Results}\label{table:logitresult}
\resizebox{\textwidth}{!}{
\begin{tabular}{lccc}
\hline
\multicolumn{1}{c}{\textbf{Variable}}                                    & \textbf{\begin{tabular}[c]{@{}c@{}}Estimated \\ Coefficient$^a$\end{tabular}}           & \textbf{\begin{tabular}[c]{@{}c@{}}t-\\ Statistic\end{tabular}}           & \textbf{\begin{tabular}[c]{@{}c@{}}Risk$^b$ \\ Ratio\end{tabular}}          \\ \hline
Constant                                                                 & -1.1129 (0.039)$^{**}$                          & -28.810                        & -1.189 (-1.037)              \\
No. of Combination Trucks in AADT                                     & -0.2095 (0.062)$^{**}$                          & -3.392                         & -0.331 (-0.088)              \\
No. of Single-Unit Trucks in AADT                                     & 0.9006 (0.069)$^{**}$                           & 12.985                         & 0.765 (1.037)                \\
Median Width of Area Separating Opposing Lanes                           & 0.0892 (0.034)$^{**}$                           & 2.651                          & 0.023 (0.155)                \\
No. of Lanes on Roadway for Continuous Travel in One Direction           & -0.4268 (0.093)$^{**}$                          & -4.580                         & -0.609 (-0.244)              \\
Percent of Single-Unit-Trucks in AADT                                    & -0.4085 (0.044)$^{**}$                          & -9.381                         & -0.494 (-0.323)              \\
Roadbed Width                                                            & -0.9389 (0.097)$^{**}$                          & -9.630                         & -1.130 (-0.748)              \\
Min. Width of Land for Road Traffic                                      & 0.2283 (0.036)$^{**}$                           & 6.324                          & 0.158 (0.299)                \\
Type of Inside Shoulder Edge: Stabilized-Surfaced with Flex              & -1.3024 (0.355)$^{**}$                          & -3.695                         & -1.993 (-0.611)              \\
Type of Outside Shoulder Edge: Stabilized-Surfaced with Flex             & 1.0803 (0.219)$^{**}$                           & 4.937                          & 0.651 (1.509)                \\ \hline 
\multicolumn{4}{l}{\begin{tabular}[c]{@{}l@{}}Log-likelihood at zero: -$-3348.9$ \\ Log-likelihood at convergence: -2941.4\\ Chi-square: 815\\ Pseudo $R^2$: 0.1217\\ $P$-value: 0.0000 \end{tabular}} \\ \hline 
\multicolumn{4}{l}{\begin{tabular}[c]{@{}l@{}}$^{**}$ \textgreater{}99\% level of significance\\ $^a$ Standard error are in parentheses\\ $^b$ Lower and upper limits at the 95\% confidence interval are in parenthese\end{tabular}}
\end{tabular}}
\end{table}

Nine road and traffic attributes significantly impact the likelihood of increasing or decreasing high-risk near-crash events at the roadway level. In particular, the number of combination trucks (heavy trucks with tailors) in the Annual Average Daily Traffic (AADT) and the percentage of single-unit trucks in AADT are associated with a reduced likelihood of high-risk events. Additionally, the number of lanes in one direction, the width of the roadbed, and the presence of an inside shoulder edge with a stabilized surface type are also correlated with a decreased probability of high-risk events. Based on the results, we find that heavy trucks may not be a primary factor in near-crash events. This could be because drivers tend to exercise greater caution when sharing the road with more heavy trucks, which may, in turn, reduce the likelihood of near-crash incidents. A higher percentage of single-unit trucks in AADT suggests a lower proportion of smaller and more maneuverable vehicles(like passenger cars) on the road. Passenger vehicles are more likely to engage in quick lane changes, sudden stops, or speeding, which can increase the risk of near-crash events. Therefore, a higher concentration of single-unit trucks could lead to fewer risky maneuvers and a safer driving environment. Moreover, the likelihood of near-crash events decreases as the number of lanes for continuous travel in one direction and the width of roadbeds increases. This suggests that when there are sufficient lanes, and the road surface is wide enough, vehicles are less likely to engage in frequent overtaking and merging, thereby further reducing the probability of near-crash occurrences. 

As reflected by their positive coefficients, the remaining four variables are positively associated with an increased likelihood of high-risk near-crash events. Specifically, the odds of the high-risk near-crash events are approximately 2.46 ($exp(0.9066)$) times higher for each unit increase in the number of single-unit trucks in AADT. Single-unit trucks, such as vans with enclosed boxes and dump trucks, are more prevalent on urban roads, where frequent stopping, starting, and lane changes occur, increasing the risk of near-crash events. The median width of the area separating opposing lanes has a smaller positive coefficient (0.0892), suggesting that wider medians slightly increase the likelihood of a near-crash event. However, the odds ratio is close to 1 ($exp(0.0892)$), indicating a marginal effect. Another factor, the minimum width of land for road traffic, has a positive coefficient of 0.2283, indicating that increased land width for road traffic slightly increases the odds of a near-crash event, with an odds ratio of 1.26($exp(0.2283)$). A possible reason for this could be that wider roads may encourage drivers to travel at higher speeds. Higher speeds reduce reaction time and increase stopping distances, elevating the likelihood of near-crash events, especially in unexpected situations \citep{millard2022width}. 

Another interesting finding is that the impacts of inside and outside road shoulder types on near-crash events are opposite. For the edge with the stabilized surface using flex, the odds of near-crash events are effectively reduced on the inside shoulder of the roadway but increase on the outside shoulder of the roadway. This difference could be attributed to these shoulders' distinct functions and associated driver behaviors. Inside shoulders, typically adjacent to the median, are less frequently used and may offer a safer, more stable refuge during emergencies, resulting in fewer near-crash events. In contrast, outside shoulders are more commonly utilized for pulling over, lane changes, or during emergencies. The stabilized surface on these shoulders might inadvertently encourage riskier behaviors, such as driving too close to the edge or using the shoulder as an additional lane, which could increase the likelihood of near-crash events. Additionally, the outside shoulder's greater exposure to environmental factors such as debris could further elevate this risk.

\subsection{The Spatial distributions of near-crash events across time periods on workdays and holidays}
After understanding the overall landscape of near-crash events in San Antonio and the factors most strongly associated with them, we grouped all events by their occurrence time to further analyze variations in their spatial distributions. Investigating the time-specific distributions of near-crash events offers deeper insights into the conditions contributing to road safety risks at specific locations and times. All detected near-crash events are grouped according to four distinct periods: Morning Peak (6AM-10AM), Daytime(10AM-4PM), Evening Peak(4PM-8PM), and Nighttime (8PM-6AM) on both workdays and holidays in November 2021. The spatial distribution of near-crash events across different periods on workdays and holidays is shown in Fig.\ref{app:crash_periods}. We adopt the $Getis$-$Ord Gi^*$ model to analyze the hot and cold spots of near-crash events in the City of San Antonio with a 1-mile fixed distance threshold calculating the spatial weights. 

\begin{figure}[htbp]
\centering
\includegraphics[width=0.95\textwidth]{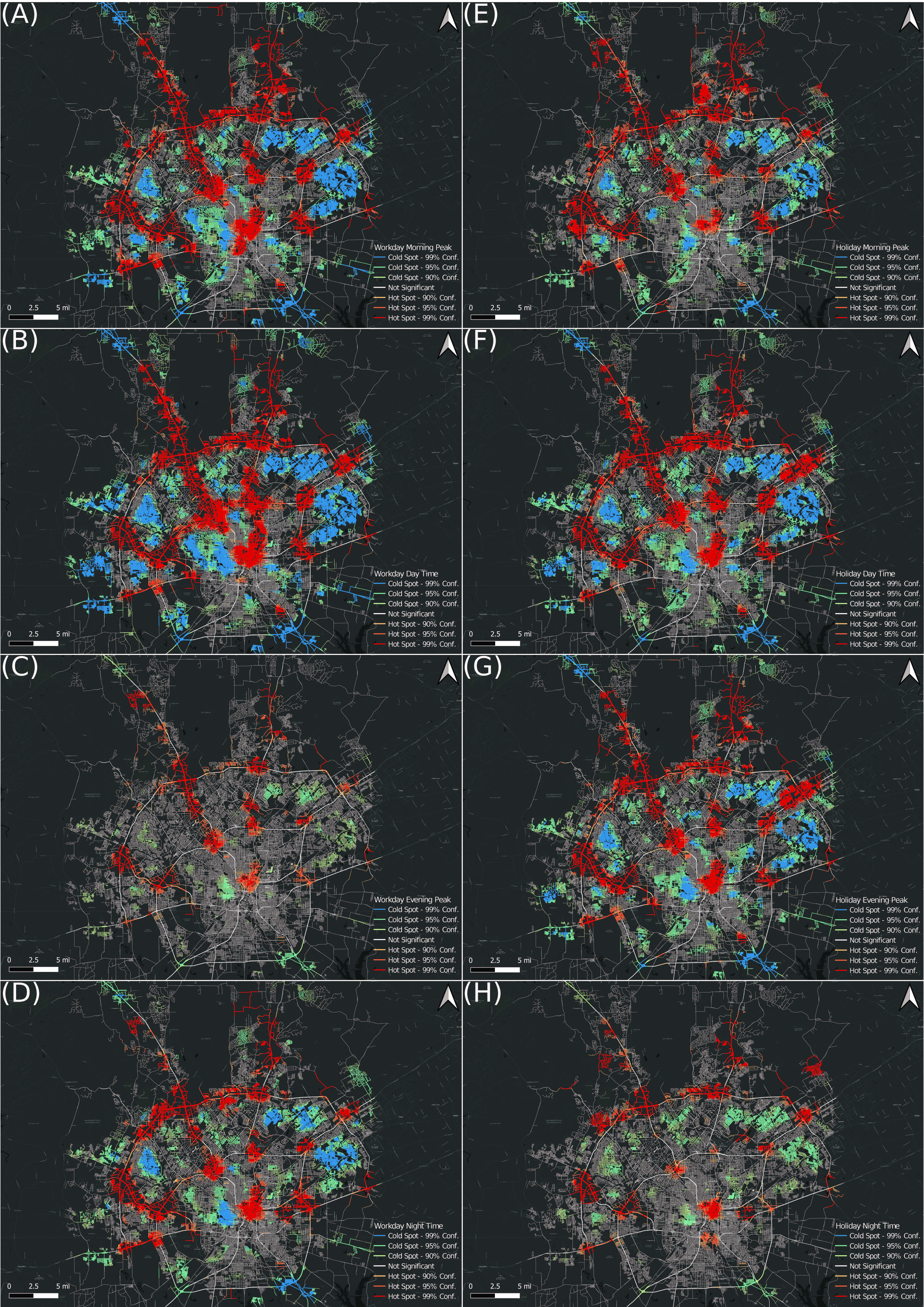}%
\caption{Spatial Distribution of Near-Crash Hot and Cold Spots by Road Segments Across Different Times of Day on Workdays and Holidays}\label{fig:road_periods}
\end{figure}

As shown in Fig.\ref{fig:road_periods}, red and blue areas indicate statistically significant hot spots where near-crash events are concentrated and cold spots with low-frequent areas, respectively. The deep red and deep blue areas with 99\% confidence demonstrate that there is only a 1\% chance that the pattern could be due to random variation. 

Fig.\ref{fig:road_periods}(A)-(D) show the high-frequent and low-frequent near-crash events across four periods in workdays, and Fig.\ref{fig:road_periods}(E)-(F) represent the near-crash patterns in holidays, respectively. Compared to spatial patterns of the holidays' morning peak events (in Fig.\ref{fig:road_periods}(A) and (E)), the high-frequent near-crash areas in the downtown of San Antonio are more significant in the workdays' morning peak. This suggests that the workday morning commute contributes significantly to the concentration of near-crash events in the downtown area, likely due to higher traffic volumes and more complex traffic flows as people travel to work. Moreover, the patterns of near-crash events in workdays' and holidays' daytime periods are similar, which indicates that daytime traffic conditions—such as steady flows of commercial vehicles, local traffic, and routine activities—remain consistent contributors to near-crash events regardless of workdays or holidays. During the evening peaks on workdays, the higher frequency of near-crash events in the downtown area and along major arteries reflects the intensity of the commute from work to home. However, the holiday evening peak in Fig.\ref{fig:road_periods}(G) shows a less intense pattern, possibly because people are engaging in a wider variety of activities, such as shopping, dining, or visiting recreational areas, which may be spread out across the city, reducing the concentration of traffic in specific areas. Furthermore, workday nighttime sees more concentrated hot spots, particularly in downtown and along key highways, driven by late-night commuting and other activities. On holiday nights, the overall risk is lower, with fewer and more dispersed hot spots, reflecting the reduced and more varied traffic flows typical of holidays.

\begin{figure}[htbp]
\centering
\includegraphics[width=0.95\textwidth]{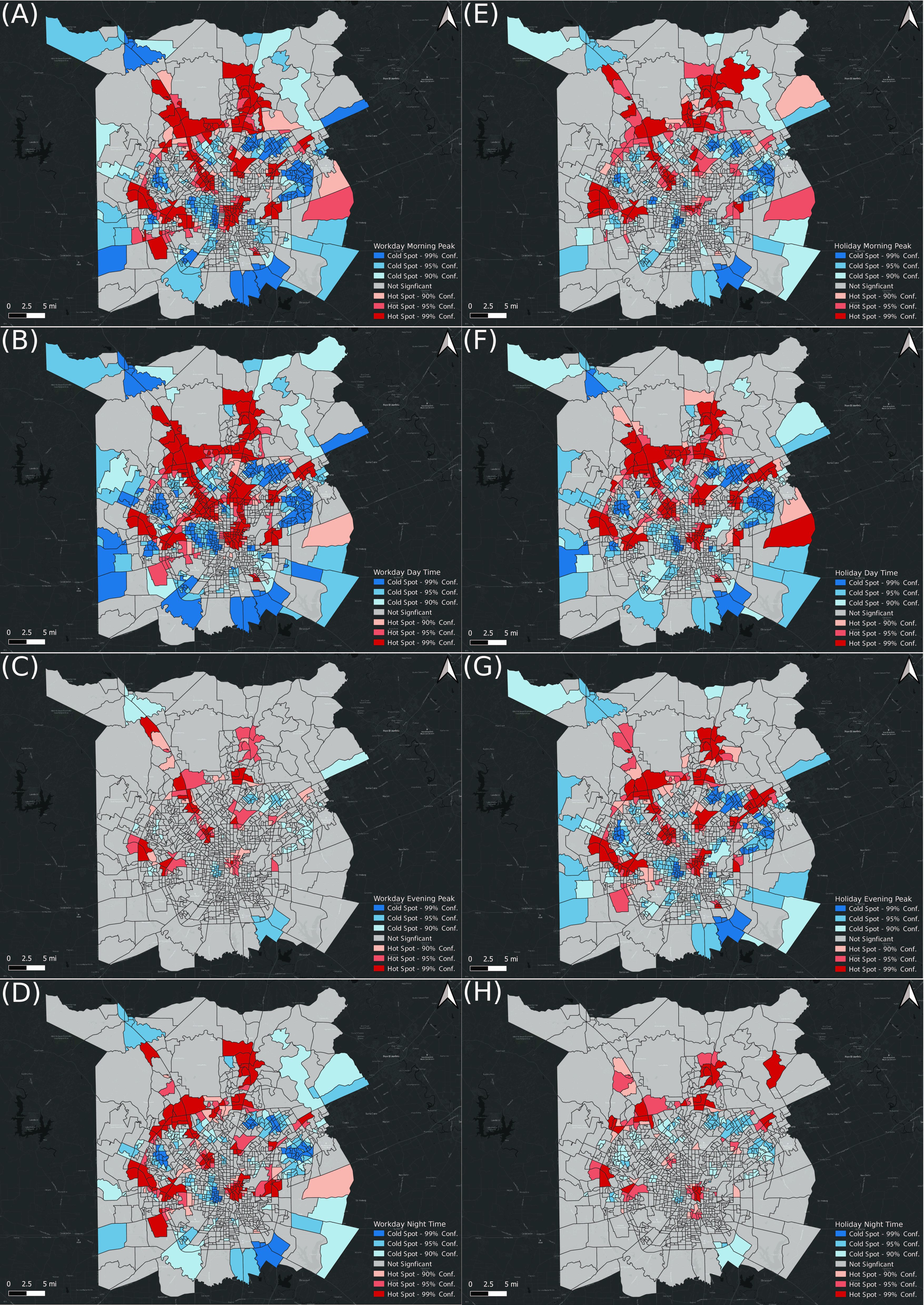}
\caption{Spatial Distribution of Near-Crash Hot and Cold Spots by Traffic Analysis Zones (TAZs) Across Different Times of Day on Workdays and Holidays}\label{fig:taz_period}
\end{figure}

Associating road-level hot spot areas with traffic analysis zones (TAZs) provides a more comprehensive understanding of the spatial distribution of traffic safety issues and their impact across broader regions. By understanding the distribution of near-crash events within TAZs, transportation planners can make more informed decisions about where to focus safety improvements, infrastructure investments, or policy changes to reduce the likelihood of crashes. As shown in Fig.\ref{fig:taz_period}, (A) through (D) show the patterns during the workday morning peak, daytime, evening peak, and nighttime periods, respectively, while (E) through (H) depict the corresponding patterns for holidays. Table \ref{table:spotsresults} demonstrates the corresponding summary of hot and cold spots of near-crash events by TAZs. 

During the workday morning peak, numerous hot spots (131 at 99\% confidence) are concentrated in downtown and northern San Antonio, reflecting the higher risk associated with the morning commute. According to the census survey, most middle- and high-income communities are clustered in the north of San Antonio, where the average number of vehicle ownership per household is more than 2, implying that there will be high-traffic zones during the workday morning peak. On holidays, although central areas still exhibit elevated near-crash risks due to ongoing holiday activities, the number of hot spots slightly decreases (121 at 99\% confidence), reflecting reduced near-crash risks. Moreover, during the daytime, the spatial patterns of hot spots change slightly between workdays and holidays, which indicates that while certain areas remain consistently high-risk due to regular activities, other areas may experience shifts in risk levels on holidays, likely due to changes in traffic pattern, such as increased recreational travel or different traffic volumes in commercial and residential zones. During the evening peak hours on holidays, the number of hot spot areas increases by 20 zones compared to the events on workdays, possibly due to different traffic patterns, such as recreational travel or evening outings. During the nighttime, near-crash hot spot areas decrease further throughout the daytime hours, especially on weekends, due to fewer nighttime trips and a lower number of nighttime events. 

\begin{table}[htbp]
\caption{Hot and Cold Spots of Near-Crash Events across Different Time Periods}\label{table:spotsresults}
\resizebox{1.\textwidth}{!}{
\begin{tabular}{cccccccccc}
\hline
\multicolumn{1}{l}{\multirow{2}{*}{}}                                          & \multirow{3}{*}{\textbf{Confidence}} & \multicolumn{4}{c}{\textbf{Workday}}                                                                                                                                                                                                                   & \multicolumn{4}{c}{\textbf{Holiday}}                                                                                                                                                                                                                  \\ \cline{3-10} 
\multicolumn{1}{l}{}                                                           &                                      & \textbf{\begin{tabular}[c]{@{}c@{}}Morning\\  Peak\end{tabular}} & \textbf{Day Time}                                       & \textbf{\begin{tabular}[c]{@{}c@{}}Evening\\ Peak\end{tabular}} & \textbf{Night Time}                                     & \textbf{\begin{tabular}[c]{@{}c@{}}Morning\\ Peak\end{tabular}} & \textbf{Day Time}                                       & \textbf{\begin{tabular}[c]{@{}c@{}}Evening\\ Peak\end{tabular}} & \textbf{Night Time}                                     \\ \hline
\multirow{5}{*}{\textbf{\begin{tabular}[c]{@{}c@{}}Cold \\ Spot\end{tabular}}} & \textbf{99\%}                        & \begin{tabular}[c]{@{}c@{}}133\\ (11.53\%)\end{tabular}          & \begin{tabular}[c]{@{}c@{}}188\\ (16.31\%)\end{tabular} & \begin{tabular}[c]{@{}c@{}}0\\ (0\%)\end{tabular}               & \begin{tabular}[c]{@{}c@{}}64\\ (5.55\%)\end{tabular}   & \begin{tabular}[c]{@{}c@{}}64\\ (5.55\%)\end{tabular}           & \begin{tabular}[c]{@{}c@{}}106\\ (9.19\%)\end{tabular}  & \begin{tabular}[c]{@{}c@{}}77\\ (6.68\%)\end{tabular}           & \begin{tabular}[c]{@{}c@{}}0\\ (0\%)\end{tabular}       \\
                                                                               & \textbf{95\%}                        & \begin{tabular}[c]{@{}c@{}}135\\ (11.71\%)\end{tabular}          & \begin{tabular}[c]{@{}c@{}}123\\ (10.67\%)\end{tabular} & \begin{tabular}[c]{@{}c@{}}26\\ (2.25\%)\end{tabular}           & \begin{tabular}[c]{@{}c@{}}101\\ (8.76\%)\end{tabular}  & \begin{tabular}[c]{@{}c@{}}117\\ (10.15\%)\end{tabular}         & \begin{tabular}[c]{@{}c@{}}131\\ (11.36\%)\end{tabular} & \begin{tabular}[c]{@{}c@{}}133\\ (11.53\%)\end{tabular}         & \begin{tabular}[c]{@{}c@{}}59\\ (5.12\%)\end{tabular}   \\
                                                                               & \textbf{90\%}                        & \begin{tabular}[c]{@{}c@{}}84\\ (7.29\%)\end{tabular}            & \begin{tabular}[c]{@{}c@{}}67\\ (5.81\%)\end{tabular}   & \begin{tabular}[c]{@{}c@{}}69\\ (5.98\%)\end{tabular}           & \begin{tabular}[c]{@{}c@{}}104\\ (9.02\%)\end{tabular}  & \begin{tabular}[c]{@{}c@{}}92\\ (7.98\%)\end{tabular}           & \begin{tabular}[c]{@{}c@{}}100\\ (8.67\%)\end{tabular}  & \begin{tabular}[c]{@{}c@{}}89\\ (7.72\%)\end{tabular}           & \begin{tabular}[c]{@{}c@{}}91\\ (7.89)\end{tabular}     \\ \hline
\multicolumn{2}{c}{\textbf{Not Significant}}                                                                          & \begin{tabular}[c]{@{}c@{}}579\\ (50.22\%)\end{tabular}          & \begin{tabular}[c]{@{}c@{}}517\\ (44.84\%)\end{tabular} & \begin{tabular}[c]{@{}c@{}}953\\ (82.65\%)\end{tabular}         & \begin{tabular}[c]{@{}c@{}}720\\ (62.45\%)\end{tabular} & \begin{tabular}[c]{@{}c@{}}710\\ (61.58\%)\end{tabular}         & \begin{tabular}[c]{@{}c@{}}611\\ (52.99\%)\end{tabular} & \begin{tabular}[c]{@{}c@{}}674\\ (58.46\%)\end{tabular}         & \begin{tabular}[c]{@{}c@{}}901\\ (78.14\%)\end{tabular} \\ \hline
\multirow{5}{*}{\textbf{\begin{tabular}[c]{@{}c@{}}Hot\\ Spot\end{tabular}}}   & \textbf{90\%}                        & \begin{tabular}[c]{@{}c@{}}38\\ (3.30\%)\end{tabular}            & \begin{tabular}[c]{@{}c@{}}32\\ (2.78\%)\end{tabular}   & \begin{tabular}[c]{@{}c@{}}38\\ (3.30\%)\end{tabular}           & \begin{tabular}[c]{@{}c@{}}37\\ (3.21\%)\end{tabular}   & \begin{tabular}[c]{@{}c@{}}34\\ (2.95\%)\end{tabular}           & \begin{tabular}[c]{@{}c@{}}33\\ (2.86\%)\end{tabular}   & \begin{tabular}[c]{@{}c@{}}38\\ (3.30\%)\end{tabular}           & \begin{tabular}[c]{@{}c@{}}31\\ (2.69\%)\end{tabular}   \\
                                                                               & \textbf{95\%}                        & \begin{tabular}[c]{@{}c@{}}53\\ (4.60\%)\end{tabular}            & \begin{tabular}[c]{@{}c@{}}52\\ (4.51\%)\end{tabular}   & \begin{tabular}[c]{@{}c@{}}41\\ (3.56\%)\end{tabular}           & \begin{tabular}[c]{@{}c@{}}42\\ (3.64\%)\end{tabular}   & \begin{tabular}[c]{@{}c@{}}65\\ (5.64\%)\end{tabular}           & \begin{tabular}[c]{@{}c@{}}51\\ (4.42\%)\end{tabular}   & \begin{tabular}[c]{@{}c@{}}48\\ (4.16\%)\end{tabular}           & \begin{tabular}[c]{@{}c@{}}43\\ (3.73\%)\end{tabular}   \\
                                                                               & \textbf{99\%}                        & \begin{tabular}[c]{@{}c@{}}131\\ (11.36\%)\end{tabular}          & \begin{tabular}[c]{@{}c@{}}174\\ (15.09\%)\end{tabular} & \begin{tabular}[c]{@{}c@{}}26\\ (2.25\%)\end{tabular}           & \begin{tabular}[c]{@{}c@{}}85\\ (7.37\%)\end{tabular}   & \begin{tabular}[c]{@{}c@{}}71\\ (6.16\%)\end{tabular}           & \begin{tabular}[c]{@{}c@{}}121\\ (10.49\%)\end{tabular} & \begin{tabular}[c]{@{}c@{}}94\\ (8.15\%)\end{tabular}           & \begin{tabular}[c]{@{}c@{}}28\\ (2.43\%)\end{tabular}   \\ \hline
\end{tabular}}
\end{table}

\subsection{The Near-crash events on road categories across workdays and holidays}
\begin{figure}[htbp]
\centering
\includegraphics[width=0.9\textwidth]{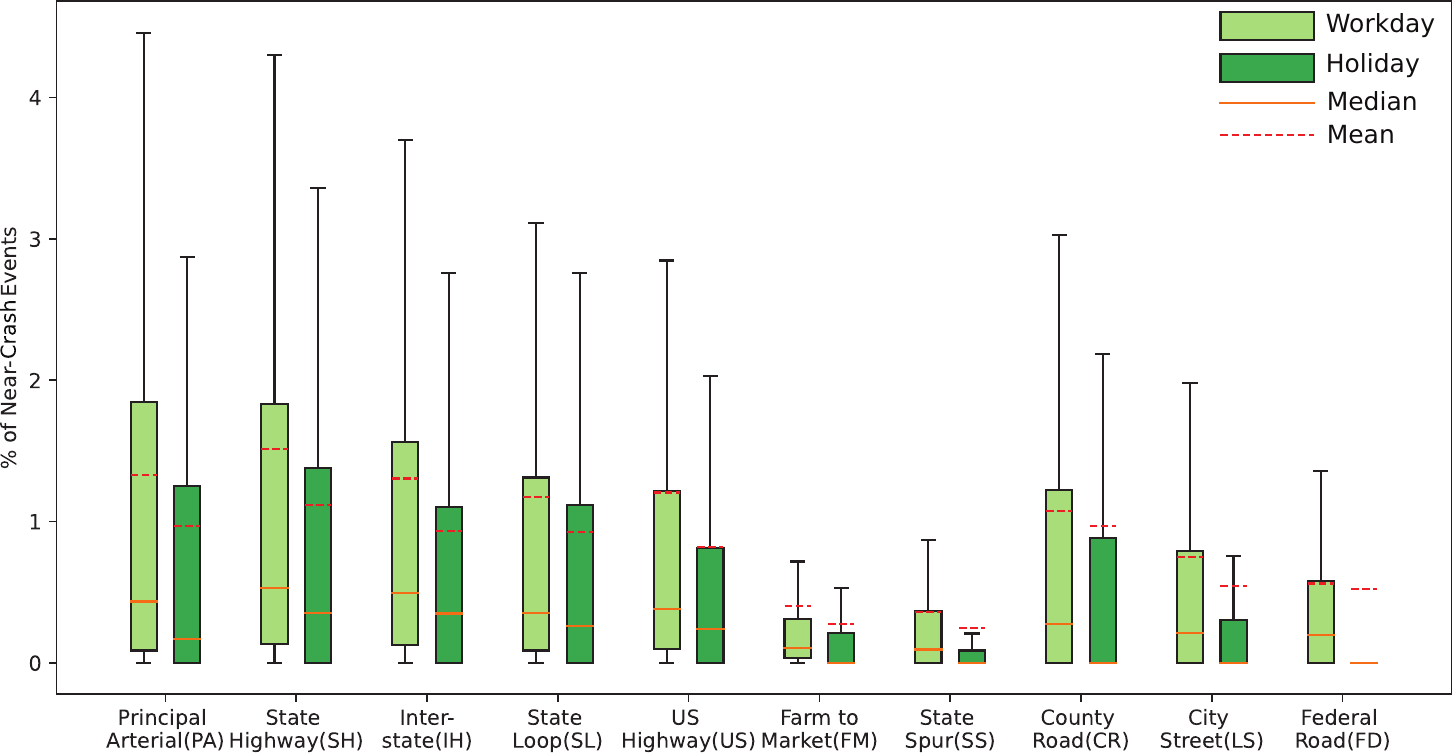}
\caption{Overall Near-Crash Risks in Road Categories on Workday and Holiday}\label{fig:category_overall}
\end{figure}

Investigating near-crash events across road categories helps inform broader transportation policy and urban planning decisions. Different road categories (e.g., highways, arterial roads, residential streets) may exhibit varying levels of risk depending on the day. It provides data-driven insights that can be used to improve road designs, manage traffic flow, and enhance overall road safety according to the unique needs of workdays versus holidays. In this study, we classify all road segments into ten categories, including Principal Arterial (PA), State Highway (SH), Inter-state (IH), State-Loop (SL), US Highway (US), Farm to Market (FM), State Spur (SS), Country Road (CR), City Local Street (LS), and Federal Road (RD). Fig.\ref{fig:category_overall} shows the overall near-crash risk in each road category on workdays and holidays. 

In general, near-crash risks are higher on workdays across most road categories, particularly on primary and heavily trafficked roads such as Principal Arterials, State Highways, and Interstates. The reduction in risk on holidays across all road categories suggests that lower traffic volumes and varied travel patterns contribute to safer driving conditions. Compared to other road categories, Principal Arterials, State Highways, and Interstates are at higher near-crash risk levels with greater mean and median and wider interquartile ranges. These three types of roads typically carry the bulk of traffic with various vehicle types and are designed for higher travel speeds. Large traffic volumes and high travel speeds will reduce the time available for drivers to react to unexpected events, increasing the chances of near-crash situations. In addition, Principal Arterials and Interstates often have complex traffic designs, including multiple lanes, frequent merging and diverging points, and high-density intersections or interchanges. This complexity increases the potential for driver error and near-crash incidents, especially during peak traffic periods on workdays. In contrast, road categories like City Street and Federal Roads show significantly lower risks, likely due to lower traffic volumes, simpler road designs, and generally lower speeds. 

\begin{figure}[htbp]
\centering
\includegraphics[width=1.0\textwidth]{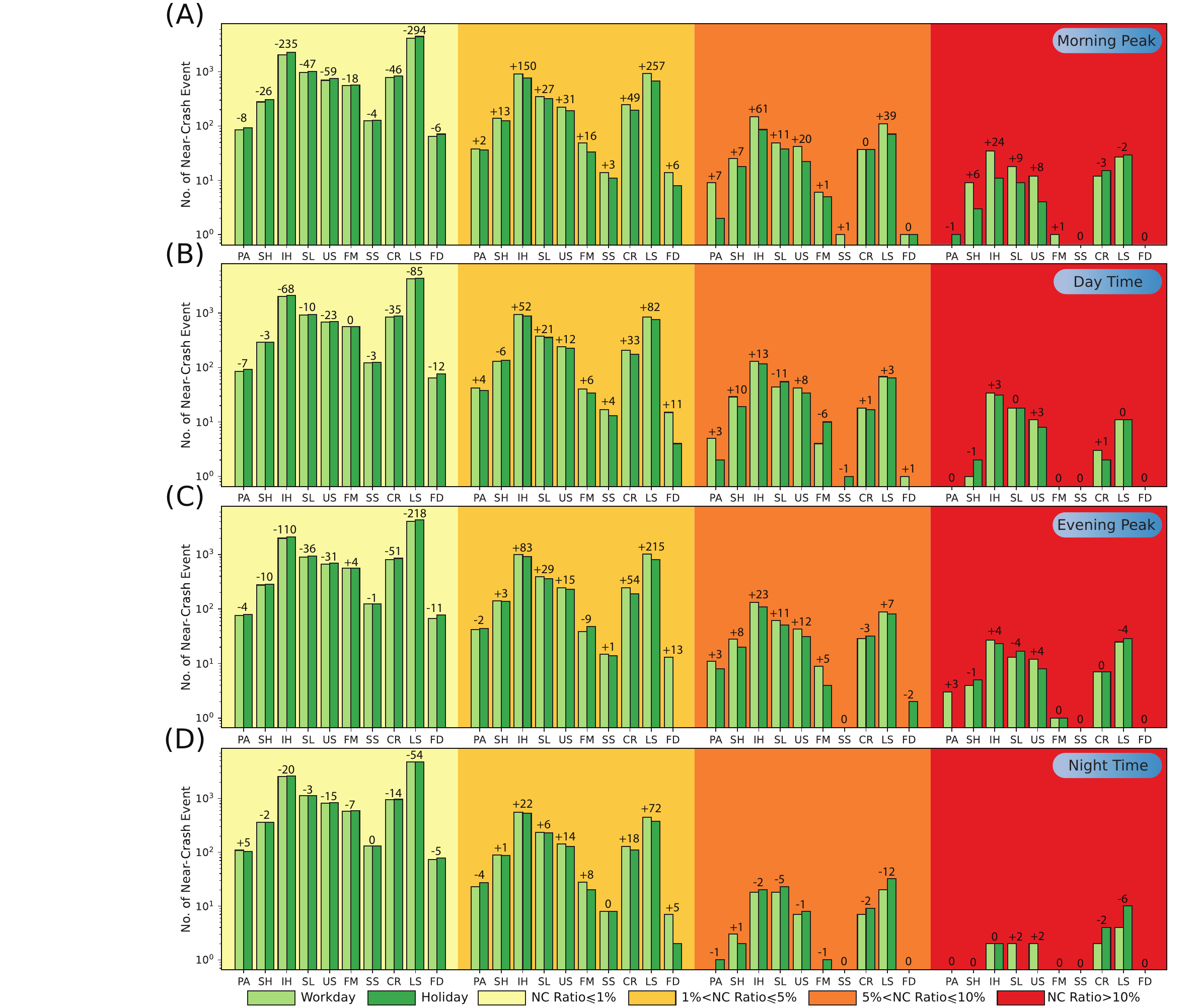}
\caption{No. of Near-Crash Events in Road Categories on Different Periods of Workday and Holiday}\label{fig:category_periods}
\end{figure}

We further analyze the distribution of near-crash events of different road categories across four periods, distinguishing between workdays and holidays. Fig.\ref{fig:category_periods} reveals that the highest number of near-crash events typically occurs on Principal Arterials and State Highways across all time periods, with noticeable increases during the morning and evening peaks. The difference between workdays and holidays is most pronounced on these road categories, indicating that commuter traffic significantly contributes to the frequency of near-crash events. Additionally, the figure highlights the near-crash event ratios across different severity thresholds, with more severe ratios (5\%-10\% and \textgreater{10\%}) being more common on PA and SH during peak hours. The analysis underscores that near-crash risks are higher on major roads and are particularly exacerbated during peak commuting times, with workdays generally observing more incidents than holidays.

\section{Discussion and Conclusion}\label{sec:conclusion}
This study leverages connected vehicle data to provide a detailed examination of near-crash events in the City of San Antonio, offering significant insights into the spatial-temporal patterns of these incidents. One of the most important findings of this study is that near-crash events are not randomly distributed but are concentrated in specific urban areas, with certain road segments exhibiting a much higher frequency of incidents. The study detected over 3.3 million near-crash events across 13,231 road segments, with nearly 76 segments identified as high-risk areas where the near-crash event ratio exceeds 10\%. Specifically, those urban areas with high traffic volumes and complex road geometries are more prone to near-crash events, such as downtown and residential areas with high vehicle ownership. Moreover, the analysis of vehicle speed and distance at the time of near-crash events underscores the significant role that driving behavior, such as following too closely and traveling at high speeds, plays in contributing to near-crash incidents. Another important finding is that the study demonstrates that the environmental and structural features of roads, such as the number of lanes, road width, and the presence of shoulder types, significantly influence the likelihood of near-crash events. These roads' environmental and structural features emphasize the need for a holistic approach to road design and maintenance that considers the capacity and efficiency of roads and their safety implications. 

Spatial-temporal variations of near-crash events across different times of day on both workdays and holidays, uncovering critical insights into how road safety risks fluctuate based on time and day. The results show that near-crash events are significantly more concentrated during peak commuting hours on workdays, particularly in the downtown area and along major thoroughfares, where traffic volumes are highest. While near-crash risks generally decrease on holidays, the spatial patterns remain consistent, especially during daytime hours, indicating that some urban regions maintain elevated risks regardless of the day. However, during the evening and nighttime periods, the differences between workdays and holidays become more pronounced, with workdays exhibiting more concentrated hot spots, particularly in central and high-traffic areas. These findings suggest that targeted interventions during specific time periods could be highly effective in mitigating near-crash events and improving overall traffic safety in San Antonio.

In addition, analyzing near-crash events across different road categories during workdays and holidays offers valuable insights into the patterns of near-crash event distribution on various roads. Near-crash risks are consistently higher on Principal Arterials, State Highways, and Interstates, particularly on workdays when traffic volumes and road use intensity are more significant. These road categories, designed for higher speeds and accommodating a diverse mix of vehicles, present increased opportunities for near-crash events, especially during peak traffic times. Also, while near-crash risks generally decrease on holidays across most road categories, the reduction is not uniform, with some roads still exhibiting significant risks. To sum up, the combination of high traffic volumes, higher speeds, and complex road geometries on these major roadways creates environments where near-crash incidents are more likely to occur.

The findings of this study have significant implications for traffic safety management, urban planning, and policy-making, particularly in rapidly growing cities like the City of San Antonio. One of the critical implications is that this study supplements existing traffic crash records and sheds light on areas with high traffic collision risks, which can spur proactive interventions to reduce the risks even before any crashes happen in some specific road segments. Implementing specific safety measures on these road segments —such as dynamic speed limits during peak hours, advanced warning systems for sudden stops, and dedicated lanes for heavy vehicles—could substantially reduce the likelihood of traffic crash incidents.

Moreover, the study’s temporal analysis highlights the importance of time-specific traffic interventions. The significant increase in near-crash events during morning and evening peaks on workdays suggests that traffic safety strategies should be adaptable to daily fluctuations in traffic conditions. For instance, increasing law enforcement presence during peak hours, optimizing traffic signal timings to manage flow more effectively, and encouraging staggered work hours to reduce congestion could help mitigate the risks associated with high traffic volumes. Additionally, the persistence of near-crash risks on holidays, particularly in areas that typically experience heavy traffic, suggests that holiday-specific traffic management plans might be necessary. These plans could include public awareness campaigns to remind drivers of safe driving practices during holidays when travel patterns may differ from regular workdays. Enhanced monitoring and temporary traffic control measures in areas known for recreational travel could also help address the unique challenges of holiday traffic.

At last, the insights gained from this study could inform the development of new technologies and data-driven tools for traffic management. The use of connected vehicle data in this study demonstrates the potential for real-time traffic monitoring systems that can identify and respond to emerging safety risks as they occur. These systems could be integrated with existing traffic management infrastructure to provide real-time alerts to drivers, adjust traffic signals dynamically, or even deploy emergency response teams more effectively during a near-crash or actual crash.

While this study provides valuable insights into the spatial-temporal patterns of near-crash events, several limitations should be acknowledged. First, the study relies on connected vehicle data from a specific period (November 2021), which may not fully capture seasonal variations or long-term trends in road safety risks. Another limitation is the potential under-representation of certain types of near-crash events, particularly those involving non-connected vehicles or occurring in areas with lower data coverage. In the future, we will address these limitations by collecting more connected vehicle data over a more extended period to explore the long-term trends in road safety risks and integrating additional data sources, such as crash reports or data from other connected vehicle platforms, to provide a more comprehensive analysis of road safety risks.

\section*{CRediT authorship contribution statement}
Xinyu Li: Formal analysis, Methodology, Visualization, Writing – original draft; Writing – review \& editing. Dayong (Jason) Wu: Data curation, Funding acquisition, Methodology, Investigation, Writing – review \& editing. Xinyue Ye: Investigation, Writing – review \& editing. Quan Sun: Methodology, Writing – review \& editing.

\section*{Declaration of competing interest}
The authors declare that they have no known competing financial interests or personal relationships that could have appeared to influence the work reported in this paper.
\section*{Data availability}
The authors do not have permission to share data.
\section*{Acknowledgements}
 This research was supported by the Texas Department of Transportation funding project under TxDOT 0-7200 (FAIN \#693JJ22330000Y560TX0511224).

\bibliographystyle{elsarticle-harv} 
\bibliography{biography}

\pagebreak

\appendix
\section{Appendix}
\label{app1}

\begin{table}[htbp]
\label{app1:table}
\caption{Traffic and Environment Characteristics for Logic Regression Model }
\resizebox{1.\textwidth}{!}{
\begin{tabular}{ccccccc}
Variable                              & Mean               & Median             & Std.               & Min                & Max                & Description                                                                                                                                                                                                                                                                                            \\\hline
AADT CURRENT                          & 67969.81           & 42740.00           & 65473.35           & 54.00              & 262595.00          & Current annual average daily traffic                                                                                                                                                                                                                                                                   \\
AADT TRAFFIC TRUCKS                   & 4743.14            & 1949.00            & 5985.56            & 0.00               & 28646.00           & Trucks in AADT                                                                                                                                                                                                                                                                                         \\
TRUCK AADT PCT                        & 6.26               & 4.40               & 5.04               & 0.00               & 32.20              & Percentage of trucks in AADT                                                                                                                                                                                                                                                                           \\
PERCENT SINGLE TRUCK AADT             & 3.27               & 2.80               & 1.64               & 0.00               & 17.10              & Percentage of single-unit trucks in AADT                                                                                                                                                                                                                                                               \\
PERCENT COMBO TRUCK AADT              & 2.98               & 1.50               & 3.85               & 0.00               & 24.50              & Percentage of combination-unit trucks in AADT                                                                                                                                                                                                                                                          \\
AADT TRAFFIC SINGLE UNIT TRUCKS       & 2219.00            & 1277.00            & 2185.49            & 0.00               & 8998.00            & No. of single-unit trucks in AADT                                                                                                                                                                                                                                                                      \\
AADT COMBINATION UNIT TRUCKS          & 2524.14            & 769.00             & 4050.51            & 0.00               & 20223.00           & No. of combination-unit trucks in AADT                                                                                                                                                                                                                                                                 \\
HPMS MEDIAN WIDTH                     & 20.25              & 11.00              & 23.03              & 0.00               & 96.00              & Median Width of Area Separating Opposing Lanes                                                                                                                                                                                                                                                         \\
NUMBER OF THROUGH LANES               & 3.17               & 2.00               & 1.80               & 1.00               & 12.00              & No. of Lanes on Roadway for Continuous Travel in One Direction                                                                                                                                                                                                                                         \\
ROADBED WIDTH                         & 52.75              & 42.00              & 30.87              & 0.00               & 236.00             & Roadbed Width                                                                                                                                                                                                                                                                                          \\
RIGHT OF WAY WIDTH MINIMUM            & 263.13             & 300.00             & 101.63             & 0.00               & 500.00             & Min. Width of Land for Road Traffic                                                                                                                                                                                                                                                                    \\\hline
SHOULDER TYPE INSIDE                 & -                  & -                  & -                  & -                  & -                  & \begin{tabular}[c]{@{}c@{}}0=None (unpaved) \\ 1=Bituminous Surface (paved) \\ 2=Concrete Surface (paved) \\ 3=Stabilized-Surfaced with Flex (unpaved) \\ 4=Combination-Surface / Stabilized (unpaved) \\ 5=Earth-with or without turf (unpaved)\\ 6=Brick \\ 99=Unknown\end{tabular} \\\hline
SHOULDER TYPE OUTSIDE                 & -                  & -                  & -                  & -                  & -                  & \begin{tabular}[c]{@{}c@{}}0=None (unpaved) \\ 1=Bituminous Surface (paved) \\ 2=Concrete Surface (paved) \\ 3=Stabilized-Surfaced with Flex (unpaved) \\ 4=Combination-Surface / Stabilized (unpaved) \\ 5=Earth-with or without turf (unpaved)\\ 6=Brick \\ 99=Unknown\end{tabular} \\\hline
SURFACE TYPE                 & -                  & -                  & -                  & -                  & -                  & \begin{tabular}[c]{@{}c@{}}1=Continuously Reinforced Concrete \\ 2=Jointed Reinforced Concrete \\ 3=Jointed Plain Concrete\\ 4=Thick Asphaltic Concrete, over 5.5 inches \\ 5=Medium Asphaltic Concrete, 2.5 - 5.5 inches \\ 6=Thin Asphaltic Concrete, under 2.5 inches \\ 7=Composite (Asphalt Surfaced Concrete) \\ 8=Widened Composite Pavement\\ 9=Overlaid and Widened Asphaltic Concrete Pavement \\ 10=Surface Treatment Pavement\\ 11=Brick \\ 12=Bladed \\ 13=Gravel \\ 99=Unknown\end{tabular} \\  \hline
PEAK PARKING                 & -                  & -                  & -                  & -                  & -                  & \begin{tabular}[c]{@{}c@{}}1=Parking allowed on one side \\ 2=Parking allowed on both sides \\ 3=No parking allowed or none available\end{tabular}   \\ \hline
WIDENING OBSTACLE                 & -                  & -                  & -                  & -                  & -                  & \begin{tabular}[c]{@{}c@{}}0=No obstacles\\ 1=Dense development\\ 2=Major transportation facilities \\ 3=Other public facilities \\ 4=Terrain restrictions\\ 5=Historic and archaeological sites \\ 6=Environmentally sensitive areas \\ 7=Parkland\end{tabular}     \\
\hline        
\end{tabular}}
\end{table}
\begin{figure}[htbp]
\centering
\includegraphics[width=0.9\textwidth]{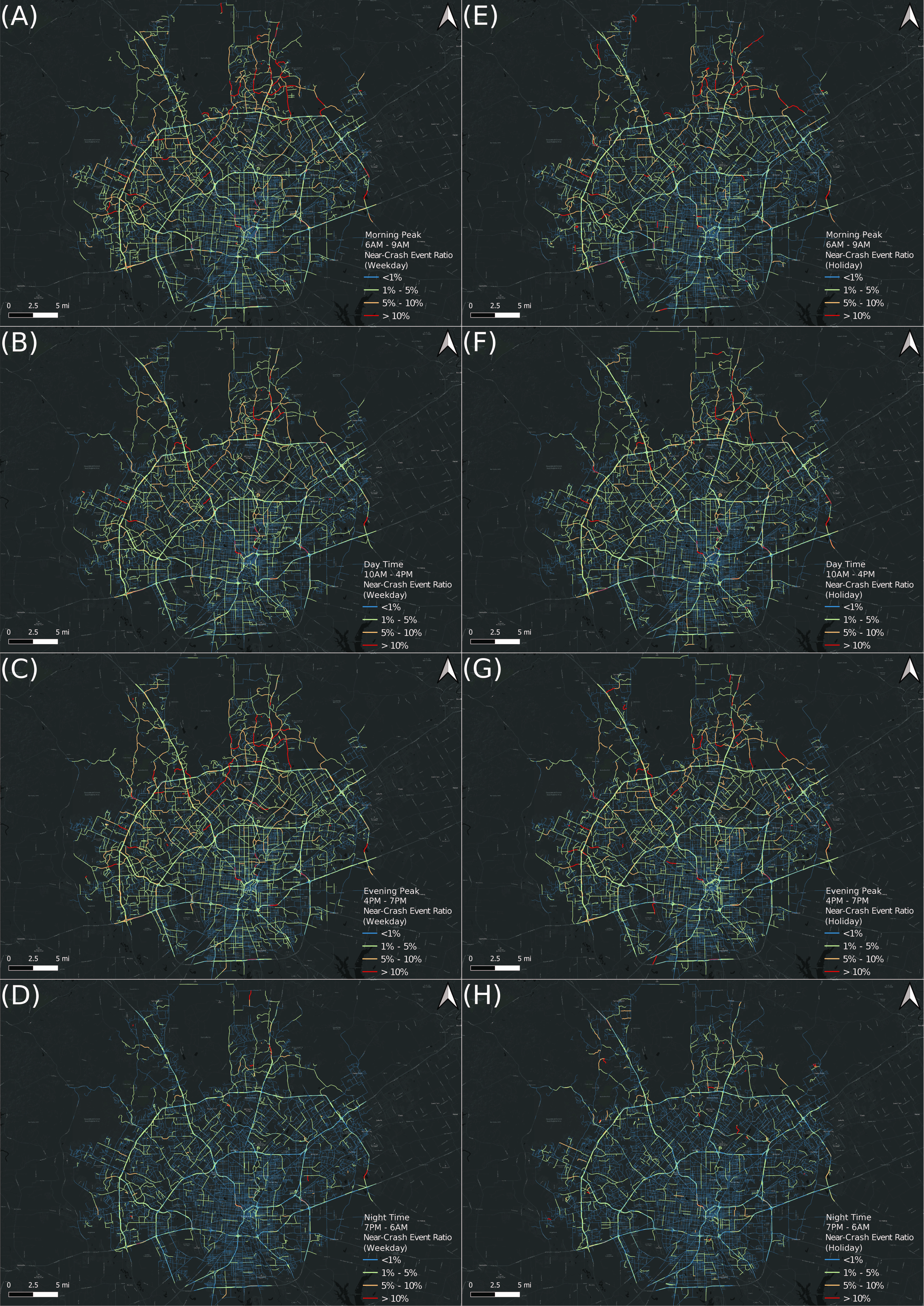}
\caption{Spatial Distribution of Near-Crash Events across Different Periods on Workdays and Holidays}\label{app:crash_periods}
\end{figure}

\end{document}